\pdfoutput=1
\documentclass[a4paper,11pt, english]{article} 
\usepackage{jheppub}
\usepackage{babel}
\usepackage{amsmath,amssymb}
\usepackage{natbib,hyperref}
\usepackage{url}
\usepackage{xspace}
\usepackage{microtype}
\usepackage{subfig}
\usepackage{graphicx}
\usepackage{textcomp}

\begin{document}
	\title{Demonstration of background rejection using deep convolutional neural networks in the NEXT experiment}
	
	\collaboration{The NEXT Collaboration}
\author[21,19]{M.~Kekic,}
\author[2]{C.~Adams,}
\author[3]{K.~Woodruff,}
\author[21,19]{J.~Renner,}
\author[20]{E.~Church,}
\author[5]{M.~Del Tutto,}
\author[21]{J.A.~Hernando~Morata,}
\author[16,9,a]{J.J.~G\'omez-Cadenas,\note[a]{NEXT Co-spokesperson.}}
\author[22]{V.~\'Alvarez,}
\author[6]{L.~Arazi,}
\author[20]{I.J.~Arnquist,}
\author[4]{C.D.R~Azevedo,}
\author[2]{K.~Bailey,}
\author[22]{F.~Ballester,}
\author[16,19]{J.M.~Benlloch-Rodr\'{i}guez,}
\author[14]{F.I.G.M.~Borges,}
\author[3]{N.~Byrnes,}
\author[19]{S.~C\'arcel,}
\author[19]{J.V.~Carri\'on,}
\author[23]{S.~Cebri\'an,}
\author[14]{C.A.N.~Conde,}
\author[11]{T.~Contreras,}
\author[21]{G.~D\'iaz,}
\author[19]{J.~D\'iaz,}
\author[5]{M.~Diesburg,}
\author[14]{J.~Escada,}
\author[22]{R.~Esteve,}
\author[6,7,19]{R.~Felkai,}
\author[13]{A.F.M.~Fernandes,}
\author[13]{L.M.P.~Fernandes,}
\author[16,9]{P.~Ferrario,}
\author[4]{A.L.~Ferreira,}
\author[13]{E.D.C.~Freitas,}
\author[16]{J.~Generowicz,}
\author[11]{S.~Ghosh,}
\author[8]{A.~Goldschmidt,}
\author[21]{D.~Gonz\'alez-D\'iaz,}
\author[11]{R.~Guenette,}
\author[10]{R.M.~Guti\'errez,}
\author[11]{J.~Haefner,}
\author[2]{K.~Hafidi,}
\author[1]{J.~Hauptman,}
\author[13]{C.A.O.~Henriques,}
\author[16]{P.~Herrero,}
\author[22]{V.~Herrero,}
\author[6,7]{Y.~Ifergan,}
\author[3]{B.J.P.~Jones,}
\author[18]{L.~Labarga,}
\author[3]{A.~Laing,}
\author[5]{P.~Lebrun,}
\author[22,19]{N.~L\'opez-March,}
\author[10]{M.~Losada,}
\author[13]{R.D.P.~Mano,}
\author[19]{J.~Mart\'in-Albo,}
\author[19,16]{A.~Mart\'inez,}
\author[19,21,b]{G.~Mart\'inez-Lema,\note[b]{Now at Weizmann Institute of Science, Israel.}}
\author[19]{M.~Mart\'inez-Vara,}
\author[3]{A.D.~McDonald,}
\author[2]{Z.-E.~Meziani}
\author[16,9]{F.~Monrabal,}
\author[13]{C.M.B.~Monteiro,}
\author[22]{F.J.~Mora,}
\author[19,16]{J.~Mu\~noz Vidal,}
\author[19]{P.~Novella,}
\author[3,a]{D.R.~Nygren,}
\author[21,19]{B.~Palmeiro,}
\author[5]{A.~Para,}
\author[12]{J.~P\'erez,}
\author[19]{M.~Querol,}
\author[6]{A.B.~Redwine,}
\author[17]{L.~Ripoll,}
\author[10]{Y.~Rodr\'iguez Garc\'ia,}
\author[22]{J.~Rodr\'iguez,}
\author[3]{L.~Rogers,}
\author[16,12]{B.~Romeo,}
\author[19]{C.~Romo-Luque,}
\author[14]{F.P.~Santos,}
\author[13]{J.M.F. dos~Santos,}
\author[6]{A.~Sim\'on,}
\author[15,c]{C.~Sofka,\note[c]{Now at University of Texas at Austin, USA.}}
\author[19]{M.~Sorel,}
\author[15]{T.~Stiegler,}
\author[22]{J.F.~Toledo,}
\author[16]{J.~Torrent,}
\author[19]{A.~Us\'on,}
\author[4]{J.F.C.A.~Veloso,}
\author[15]{R.~Webb,}
\author[6,d]{R.~Weiss-Babai,\note[d]{On leave from Soreq Nuclear Research Center, Yavneh, Israel.}}
\author[15,e]{J.T.~White,\note[e]{Deceased.}}
\author[19]{N.~Yahlali}
\emailAdd{marija.kekic@usc.es}
\affiliation[1]{
Department of Physics and Astronomy, Iowa State University, 12 Physics Hall, Ames, IA 50011-3160, USA}
\affiliation[2]{
Argonne National Laboratory, Argonne, IL 60439, USA}
\affiliation[3]{
Department of Physics, University of Texas at Arlington, Arlington, TX 76019, USA}
\affiliation[4]{
Institute of Nanostructures, Nanomodelling and Nanofabrication (i3N), Universidade de Aveiro, Campus de Santiago, Aveiro, 3810-193, Portugal}
\affiliation[5]{
Fermi National Accelerator Laboratory, Batavia, IL 60510, USA}
\affiliation[6]{
Nuclear Engineering Unit, Faculty of Engineering Sciences, Ben-Gurion University of the Negev, P.O.B. 653, Beer-Sheva, 8410501, Israel}
\affiliation[7]{
Nuclear Research Center Negev, Beer-Sheva, 84190, Israel}
\affiliation[8]{
Lawrence Berkeley National Laboratory (LBNL), 1 Cyclotron Road, Berkeley, CA 94720, USA}
\affiliation[9]{
Ikerbasque, Basque Foundation for Science, Bilbao, E-48013, Spain}
\affiliation[10]{
Centro de Investigaci\'on en Ciencias B\'asicas y Aplicadas, Universidad Antonio Nari\~no, Sede Circunvalar, Carretera 3 Este No.\ 47 A-15, Bogot\'a, Colombia}
\affiliation[11]{
Department of Physics, Harvard University, Cambridge, MA 02138, USA}
\affiliation[12]{
Laboratorio Subterr\'aneo de Canfranc, Paseo de los Ayerbe s/n, Canfranc Estaci\'on, E-22880, Spain}
\affiliation[13]{
LIBPhys, Physics Department, University of Coimbra, Rua Larga, Coimbra, 3004-516, Portugal}
\affiliation[14]{
LIP, Department of Physics, University of Coimbra, Coimbra, 3004-516, Portugal}
\affiliation[15]{
Department of Physics and Astronomy, Texas A\&M University, College Station, TX 77843-4242, USA}
\affiliation[16]{
Donostia International Physics Center (DIPC), Paseo Manuel Lardizabal, 4, Donostia-San Sebastian, E-20018, Spain}
\affiliation[17]{
Escola Polit\`ecnica Superior, Universitat de Girona, Av.~Montilivi, s/n, Girona, E-17071, Spain}
\affiliation[18]{
Departamento de F\'isica Te\'orica, Universidad Aut\'onoma de Madrid, Campus de Cantoblanco, Madrid, E-28049, Spain}
\affiliation[19]{
Instituto de F\'isica Corpuscular (IFIC), CSIC \& Universitat de Val\`encia, Calle Catedr\'atico Jos\'e Beltr\'an, 2, Paterna, E-46980, Spain}
\affiliation[20]{
Pacific Northwest National Laboratory (PNNL), Richland, WA 99352, USA}
\affiliation[21]{
Instituto Gallego de F\'isica de Altas Energ\'ias, Univ.\ de Santiago de Compostela, Campus sur, R\'ua Xos\'e Mar\'ia Su\'arez N\'u\~nez, s/n, Santiago de Compostela, E-15782, Spain}
\affiliation[22]{
Instituto de Instrumentaci\'on para Imagen Molecular (I3M), Centro Mixto CSIC - Universitat Polit\`ecnica de Val\`encia, Camino de Vera s/n, Valencia, E-46022, Spain}
\affiliation[23]{
Centro de Astropart\'iculas y F\'isica de Altas Energ\'ias (CAPA), Universidad de Zaragoza, Calle Pedro Cerbuna, 12, Zaragoza, E-50009, Spain}

	\abstract{Convolutional neural networks (CNNs) are widely used state-of-the-art computer vision tools that are becoming increasingly popular in high-energy physics. In this paper, we attempt to understand the potential of CNNs for event classification in the NEXT experiment, which will search for neutrinoless double-beta decay in $^{136}$Xe. To do so, we demonstrate the usage of CNNs for the identification of electron-positron pair production events, which exhibit a topology similar to that of a neutrinoless double-beta decay event. These events were produced in the NEXT-White high-pressure xenon TPC using 2.6-MeV gamma rays from a $^{228}$Th calibration source.  We train a network on Monte Carlo-simulated events and show that, by applying on-the-fly data augmentation, the network can be made robust against differences between simulation and data. The use of CNNs offers significant improvement in signal efficiency and background rejection when compared to previous non-CNN-based analyses.}
	
	
	\collaboration{\includegraphics[height=9mm]{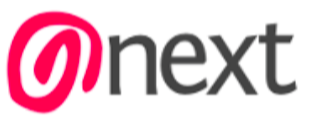}\\[6pt]
		NEXT collaboration}
	
	\maketitle
	\flushbottom

	\section{Introduction}
	Machine learning techniques have recently captured the interest of researchers in various scientific fields, including particle physics, and are now being employed in search of improved solutions to a variety of problems. In this study, we show that deep convolutional neural networks (CNNs) trained on Monte Carlo simulation can be used to classify, to a high degree of accuracy, events containing particular topologies of ionization tracks acquired from a high-pressure xenon (HPXe) time projection chamber (TPC). As CNNs trained on simulation are known to be difficult to apply directly to data due to the challenges associated with producing a Monte Carlo that perfectly matches experiment, we also present methods for extending the domain of application of a CNN trained on simulated events to include real events. We claim that our use of these methods in adapting CNNs to the experimental domain and verifying their performance is novel to the use of CNNs in the field.

Event classification is of critical importance in experiments searching for rare physics, as the successful rejection of background events can lead to significant improvements in overall sensitivity. The NEXT (Neutrino Experiment with a Xenon TPC) experiment is searching for neutrinoless double-beta decay ($0\nu\beta\beta$) in $^{136}$Xe at the Laboratorio Subterr\'aneo de Canfranc (LSC) in Spain. In the ongoing first phase of the experiment, the 5-kg-scale TPC NEXT-White \cite{Monrabal:2018xlr} has demonstrated excellent energy resolution \cite{Renner:2019csth} and the ability to reconstruct high-energy (of the order of 2~MeV) ionization tracks and distinguish between the topological signatures of two-electron and one-electron tracks \cite{Ferrario:2019}. It has also been used to perform a detailed measurement of the background distribution and is expected to be capable of measuring the $2\nu\beta\beta$ mode in $^{136}$Xe with 3.5$\sigma$ sensitivity after 1 year of data-taking \cite{Novella:2019}. The next phase of the experiment, the 100-kg-scale detector NEXT-100, will search for the $0\nu\beta\beta$ mode at $Q_{\beta\beta}$, around 2.5~MeV. New techniques such as CNNs, which analyze the topology of an event near $Q_{\beta\beta}$ and aim to eliminate background events, are becoming more relevant and essential to reaching the best possible sensitivity.

Machine learning techniques have seen many recent applications in physics \cite{Giuseppe:2019}. In neutrino physics in particular, CNNs have been applied to particle identification in sampling calorimeters in the NOvA experiment \cite{Aurisano:2016}. The MicroBooNE experiment has also employed CNNs for event classification and localization \cite{Acciarri:2017} and track segmentation \cite{Adams:2019} in liquid argon TPCs. IceCube has applied graph neural networks to perform neutrino event classification \cite{Choma:2018}, and DayaBay identified antineutrino events in gadolinium-doped liquid scintillator detectors using CNNs and convolutional autoencoders \cite{Racah:2016}. Experiments searching for $0\nu\beta\beta$ decay have also employed CNNs: EXO has studied the use of CNNs to extract event energy and position from raw-waveform information in a liquid xenon TPC \cite{Delaquis:2018} and PandaX-III has performed simulation studies demonstrating the use of CNNs for background rejection in a HPXe gas TPC with a Micromegas-based readout \cite{Qiao:2018}. Further simulation studies in HPXe TPCs with a charge readout scheme (``Topmetal'') allowing for detailed 3D track reconstruction have also shown the potential of CNNs for background rejection in $0\nu\beta\beta$ searches \cite{Ai:2018}. NEXT has also presented an initial simulation study \cite{Renner:2017ey} of the use of CNNs for background rejection. In this study we show that CNNs can be applied to real NEXT data, using electron-positron pair production to generate events with a two-electron ``$\beta\beta$-like'' topology and studying how the energy distribution of such events changes when varying an acceptance cut on the classification prediction of a CNN.

The paper is organized as follows: section~\ref{s.top_signature} describes the topological signature of a signal event. In section~\ref{s.data_reconstruction} the data acquisition and reconstruction is explained. A description of the CNN and training procedure, as well as evaluation on MC and data is given in section~\ref{s.dnn_analysis}. Finally, conclusions are drawn in section~\ref{s.conclusions}.

	\section{Topological signature}\label{s.top_signature}
\begin{figure}[!htb]
	\centering
	\includegraphics[width= 0.45\textwidth]{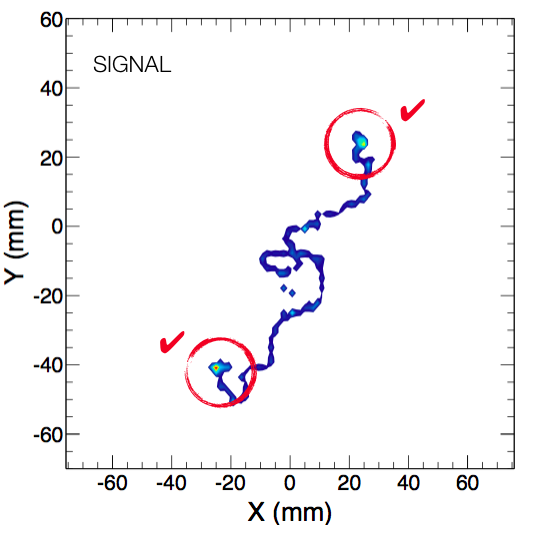}	\includegraphics[width= 0.45\textwidth]{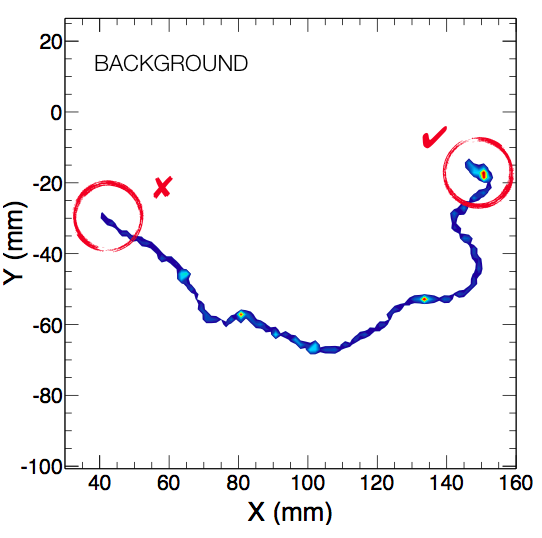}
	\caption{Energy depositions from trajectories in a Monte Carlo simulation of a $0\nu\beta\beta$ event, showing its distinct two-electron topological signature (left) compared with that of a single-electron event (right) of the same energy (figure from~\cite{Martin-Albo:2015rhw}).} \label{fig.bb0nu_signature}
\end{figure}

	In a fully-contained $0\nu\beta\beta$ event recorded by a HPXe TPC, two energetic electrons produce ionization tracks emanating from a common vertex. Though the fraction of energy $Q_{\beta\beta}$ carried by each individual electron may differ event-by-event, the general pattern observed is similar for the majority of events, and consists of an extended track capped on both ends by two ``blobs'', or regions of relatively higher ionization density. These regions are present due to the increase in stopping power experienced by electrons in xenon gas as they slow to lower energies. They provide a distinct signature for $0\nu\beta\beta$ decay, as measured tracks with similar energy produced by single electrons,\footnote{Events with multiple tracks are easier to reject simply by counting the number of isolated depositions.} for example, photoelectric interactions of background gamma radiation, contain only one such ``blob''. The use of this signature, illustrated in Fig.~\ref{fig.bb0nu_signature}, in performing background rejection is an essential part of the NEXT approach to maximizing sensitivity to $0\nu\beta\beta$ decay.

In order to demonstrate this approach experimentally, a reliable source of events with a similar topological signature is necessary. Electron-positron pair production by high energy gammas, followed by the subsequent escape from the active volume of the two 511-keV gamma rays produced in positron annihilation (``double-escape''), leaves a two-blob track formed by the electron and positron emitted from a common vertex, similar to the track that would be left by a $0\nu\beta\beta$ event. In this study, we use gamma rays of energy 2614.5~keV from $^{208}$Tl (provided by a $^{228}$Th calibration source, see Fig.~\ref{fig.tpcconfig}) and observe the events in the double-escape peak at 1592~keV. This peak lies on top of a continuous background of single-electron tracks from Compton scattering of the calibration gamma rays and other background radiation. Experimentally, then, we have a sample containing $0\nu\beta\beta$-like events and background-like events. By evaluating these events with a Monte-Carlo-trained neural network and studying the resulting distribution of accepted events, we can demonstrate, using real data acquired with the NEXT-White TPC, the potential performance of such a network when employed in a $0\nu\beta\beta$ search. These results can be compared to a similar, non-CNN-based analysis published in~\cite{Ferrario:2019}.

	\section{Data acquisition and analysis}\label{s.data_reconstruction}
	\subsection{The NEXT-White TPC}
\begin{figure}[!htb]
	\centering
	\includegraphics[width= 0.75\textwidth]{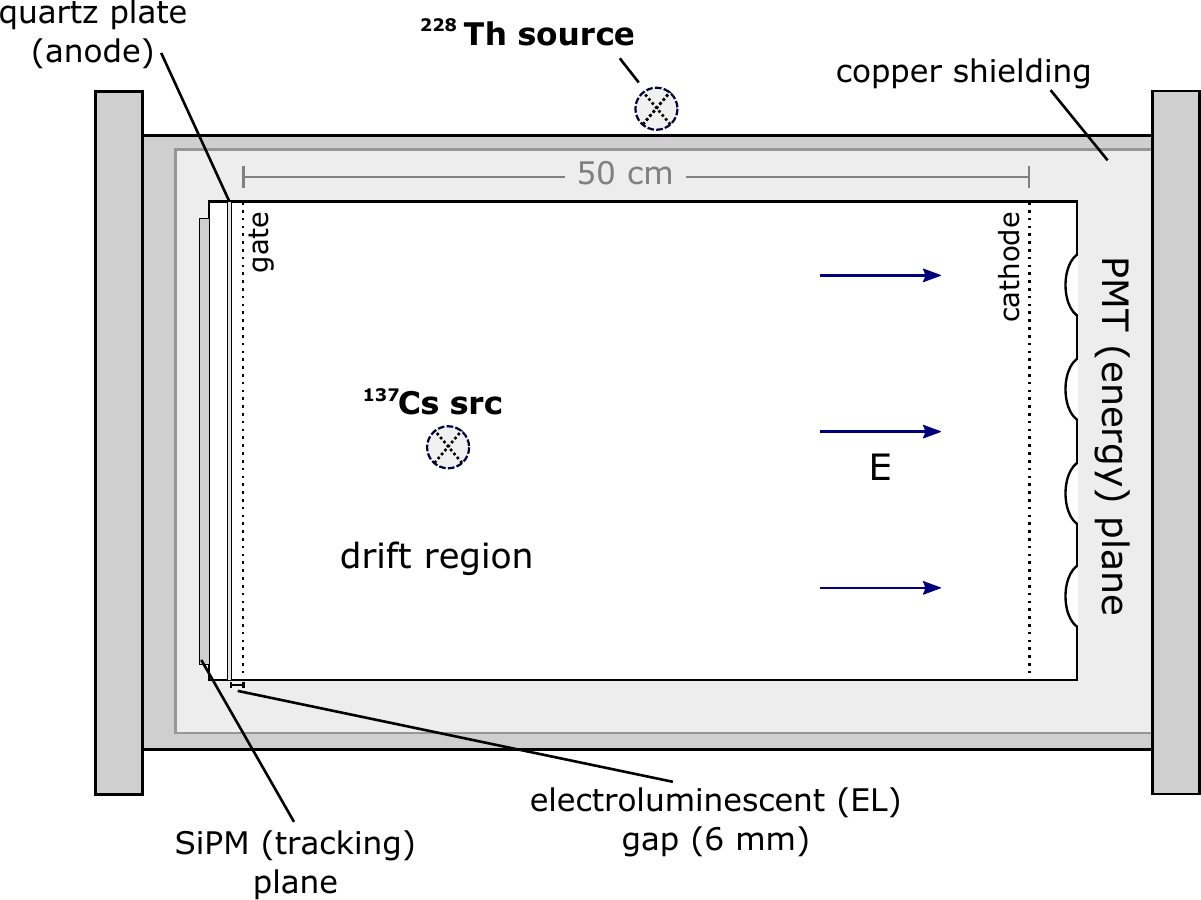}
	\caption{Schematic of the NEXT-White TPC, showing the positioning of the calibration sources ($^{137}$Cs and $^{228}$Th) present during data acquisition for this study 
		(figure derived from~\cite{Renner:2019csth}).} \label{fig.tpcconfig}
\end{figure}

The NEXT-White TPC measures both the primary scintillation and ionization produced by a charged particle traversing its active volume of high-pressure xenon gas. The main detector components are housed in a cylindrical stainless steel pressure vessel lined with copper shielding and include two planes of photosensors, one at each end, and several semi-transparent wire meshes to which voltages are applied, defining key regions of the detector (see Fig.~\ref{fig.tpcconfig}). The two planes of photosensors are organized into an \emph{energy plane}, containing 12~PMTs (photomultiplier tubes, Hamamatsu model R11410-10) behind the cathode, and a \emph{tracking plane} containing a grid of 1792~SiPMs (silicon photomultipliers, SensL series-C, spaced at a 10-mm pitch) behind the anode. These sensors observe the scintillation produced in the active volume of the detector by ionizing radiation, including primary scintillation produced by excitations of the xenon atoms during the creation of the ionization track and secondary scintillation produced by \emph{electroluminescence} (EL) of the ionization electrons. Note that in practice only the PMTs observe a consistently measurable primary scintillation signal, while EL is observed by both the PMTs and the SiPMs.

EL occurs after the electrons of the ionization track are drifted through the active region by an electric field (of order 400~V/cm) created by application of high voltage to the cathode ($-30$~kV) and gate ($-7.6$~keV) meshes and arrive at the EL gap, a narrow (6~mm) region defined by the gate mesh and a grounded quartz plate on which a conductive indium tin oxide (ITO) coating has been deposited. The large voltage drop over the narrow gap between the gate and the grounded plate creates an electric field high enough to accelerate the electrons to energies sufficient to excite the xenon without producing further ionization, allowing for better energy resolution compared to charge-avalanche detectors~\cite{Nygren:2009zz}. The subsequent decay of these excitations leads to EL scintillation, yielding of the order 500--1000 photons per electron traversing the EL gap. These photons, produced just in front of the tracking plane, cast a pattern of light on the SiPMs that can be used to reconstruct the $(x,y)$ location of the ionization. The PMTs located in the energy plane on the opposite side of the detector see a more uniform distribution of light, including EL photons that have undergone a number of reflections in the detector, and record a greater total number of photons for a more precise measurement of the energy. The time difference between the observation of the primary scintillation (called S1) and secondary EL scintillation (called S2) gives the distance drifted by the ionization electrons before arriving at the EL region, corresponding to the $z$ location at which this ionization was produced.

\subsection{Event reconstruction}\label{ss.evt_reco}
The data used in this study consisted of events with total energy near 1.6~MeV, including electron-positron events produced in pair production interactions from a 2.6-MeV gamma ray (see section~\ref{s.top_signature}) and background events, mostly due to Compton scattering of the same 2.6-MeV gamma rays.\footnote{Environmental radioactivity was negligible compared to that of the source.} The acquired signals for each event consisted of 12~PMT waveforms sampled at 25-ns intervals and 1792~SiPM waveforms sampled at 1-\textmu s intervals for a total duration per read-out greater than the TPC maximum drift (approximately 500 microseconds). The ADC counts per unit of time in each waveform were converted to photoelectrons per unit time via conversion factors established by periodic calibration using LEDs installed inside the detector, a standard procedure in NEXT-White operation. The calibrations were performed by driving LEDs installed inside the vessel with short pulses and measuring the integrated ADC counts corresponding to a single photoelectron (pe). 

The analysis of the acquired data was similar to that of~\cite{Ferrario:2019}. The 12~PMT waveforms were summed, weighted by their calibrated gains, to produce a single waveform in which scintillation pulses were identified and classified as S1 or S2 according to their shape and location within the waveform. Events containing a single S1 pulse and at least one S2 pulse were selected, and for these events, the S2 information was used to reconstruct the ionization track. To do this, the S2 information was integrated into time bins of width 2~\textmu s in both the PMTs and SiPMs. Note that to eliminate dark noise, SiPM samples with less than 1~pe were not included in the integration. 

For each time bin, one or more energy depositions (or ``\emph{hits}'') were reconstructed, and the pattern of signals observed on the SiPMs was used to determine the number of hits for a specific time bin and their corresponding $(x,y)$ coordinates. A hit was assigned to the location of all SiPMs with an observed signal greater than a given threshold, and the total energy measured by the PMTs in that time bin was redistributed among the hits according to their relative SiPM signals. 

The energy of each hit as measured by the PMTs was then corrected, hit-by-hit, by two multiplicative factors, one accounting for geometric variations in the light response in the EL plane and the other for electron attachment due to a finite electron lifetime in the gas. These correction factors were mapped out over the active volume by simultaneously acquiring events from decays of $^{83\mathrm{m}}$Kr, which was injected into the xenon gas and provided uniformly distributed point-like depositions of energy 41.5~keV~\cite{Martinez-Lema:2018ibw}. 
The $z$-coordinate of each hit in the time bin was obtained from the time difference between S1 and S2 pulses, assuming an electron drift velocity of 0.91~mm/\textmu s, as extracted from an analysis of the $^{83\mathrm{m}}$Kr events. A residual dependence of the event energy on the length of the event along the z-axis is observed, and a linear correction is performed to model this effect, which is not observed in simulation and remains to be fully understood. For details on this ``axial length'' effect, see~\cite{Renner:2019csth}.

\begin{figure}[!htb]
	\centering
	\includegraphics[width= 0.85\textwidth]{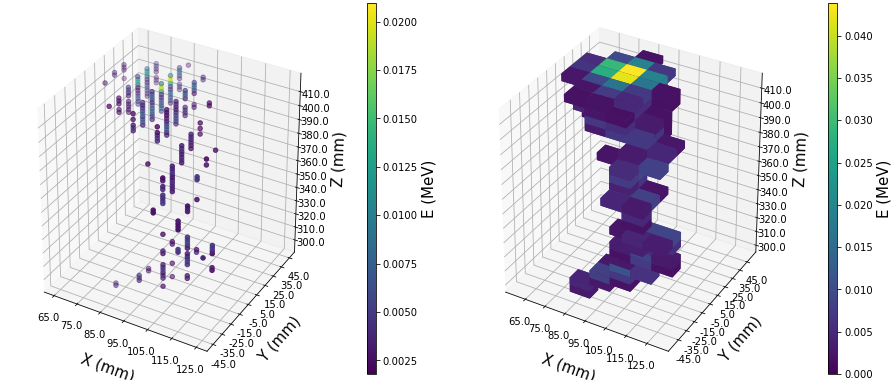}
	\caption{Reconstructed hits (left) and voxels (right) of a background Monte Carlo event. The volume within a tight bounding box encompassing the reconstructed hits is divided into $10\times10\times5\mathrm{~mm}^3$ voxels to produced the voxelized track.} \label{fig.bck_voxel}
\end{figure}

The detector volume surrounding the reconstructed hits was then partitioned into 3D voxels of side length $10\times10\times5\mathrm{~mm}^3$, and the energy of all hits that fell within each voxel was integrated. The X and Y dimensions of the individual voxels were chosen based on the 1-cm SiPM pitch, while the Z dimension was chosen to account for most of the longitudinal diffusion ($1 \sigma$ spread at maximum drift length is $\sim 2$~mm). The final voxelized track could then be considered in the neural-network-based topological analysis (see Fig.~\ref{fig.bck_voxel}).

	\section{Convolutional neural network analysis}\label{s.dnn_analysis}
	\subsection{Data preparation}\label{ss.data_prep}
To generate the events used in training the neural network, a full Monte Carlo (MC) of the detector, including the pressure vessel, internal copper shielding, and sensor planes, was constructed using Nexus~\cite{Martin-Albo:2015dza}, a simulation package for NEXT based on Geant4~\cite{Agostinelli:2002hh} (version \texttt{geant4.10.02.p01}). The $^{208}$Th calibration source decay and the resulting interactions of the decay products were simulated by Geant4, up to and including the production of the ionization track. Events in the energy range of 1.4--1.8~MeV were selected, and the subsequent electron drift, diffusion, electroluminescence, photon detection, and electronic readout processes were simulated outside of Geant4 to produce for each event a set of sensor waveforms corresponding to those acquired in NEXT-White. The analysis of data waveforms described in section~\ref{ss.evt_reco} could then be applied to these MC waveforms to produce voxelized tracks (see Fig.~\ref{fig.bck_voxel}). MC events that were fully contained in the active detector volume were used in the training set. To ensure the classification was done only based on the track topology, the energy of each voxel was scaled by the total event energy (the sum of voxel intensities for a given event was normalized to 1) such that the training data did not contain event energy information. Those events containing an electron and a positron registered in the MC true information, with no additional energy deposited by the two 511-keV gamma rays produced upon annihilation of the positron (i.e. a true ``double-escape''), were tagged as ``signal'' events and all others were tagged as ``background''. 

\begin{figure}[!tbp]
	\centering
	\begin{minipage}{0.49\textwidth}
		\centering
		\includegraphics[width=0.95\textwidth]{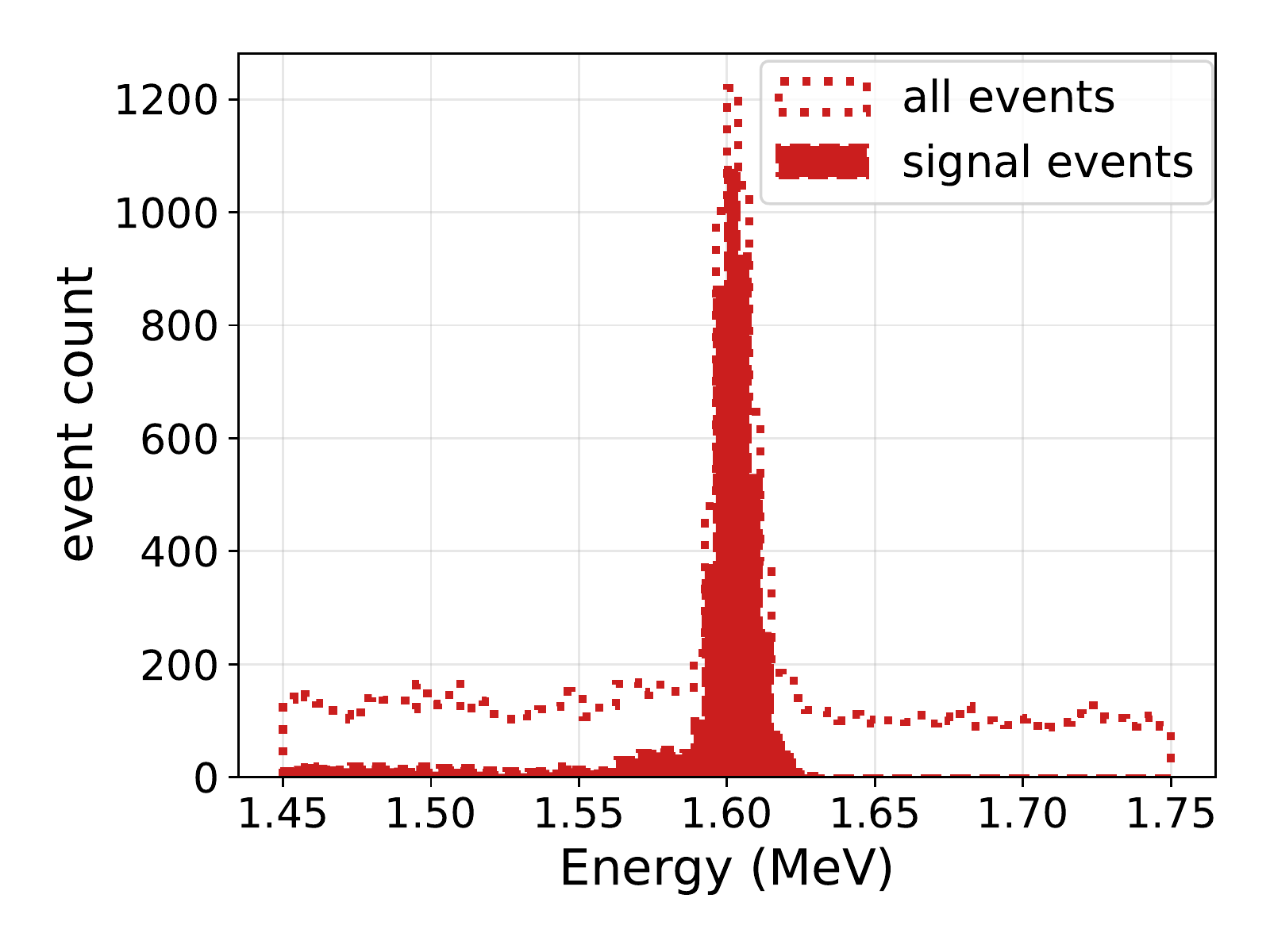}
	\end{minipage}
	\begin{minipage}{0.49\textwidth}
		\includegraphics[width=0.95\textwidth]{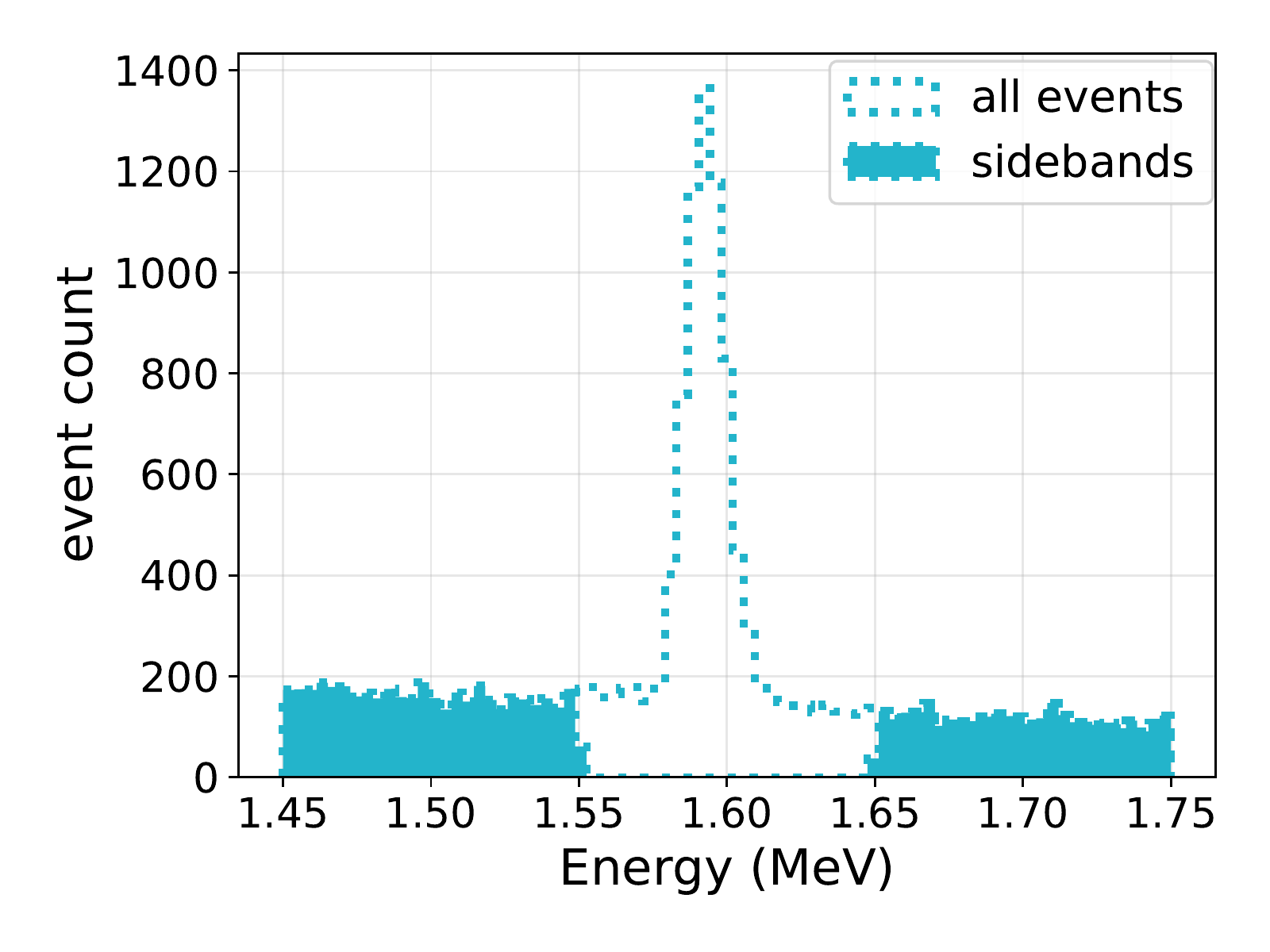}
	\end{minipage}
	\caption{Left: Energy distribution of all MC events (dashed line histogram) and of chosen signal events (solid histogram). Right: Energy distribution of experimental data events showing selected sideband events. The sidebands are 100 keV in width, with each band starting 45~keV from either side of the double escape peak. The same procedure is also used to select the sidebands in MC.}\label{fig.spectrum}
\end{figure}

In~\cite{Ferrario:2019}, an additional single-track selection cut is made, and for a fair comparison with this previous result we also apply the same cut (obtained from the standard track reconstruction, for details see~\cite{Ferrario:2019}) on test data only, for both MC and experimental data. As a reference, inside the peak energy range, the efficiency of the single-track cut was $\sim$~0.9 for signal events and $\sim$ 0.7 for background events. For signal events, additional tracks appear either from physical processes such as bremsstrahlung, or from artificial splitting of the track due to imperfect reconstruction.
The energy distribution of MC events with labeled signal events used for testing after the fiducial and single track selection cut is given in Fig.~\ref{fig.spectrum}. A continuum of signal-like events below the 1.59~MeV peak energy belongs to gammas that Compton scatter before entering the detector, and undergo a pair conversion inside the detector active volume. The topology of those events is indistinguishable from the peak events since they only differ in the absolute reconstructed energy that is not taken into account during the training, and hence they are labeled as signal.

When applying a network trained on events from one domain (MC) to events from a different domain (data), the performance will depend on the similarity between those two domains. Known differences between MC and data in high-level variables, such as track length, have been observed in a previous topological analysis~\cite{Ferrario:2019}, calling into question the performance of the MC-trained CNN when applied to data. In this study, the classification task is focused on the double-escape peak, which is clearly visible and for which we understand the underlying physical process (pair production). In this case, we could attempt to design our network to obtain optimal classification results on the acquired data and in the energy range of interest (a method to evaluate the network on double-escape data is explained in section~\ref{evaluation_data}), but we could not argue that the same procedure would work in a $0\nu\beta\beta$ search for which we do not have a confirmed understanding of the underlying physics, nor would it be justified to make predictions on the same events used in optimizing the network.

Therefore, we develop a general paradigm (as described in section~\ref{training_proc}) that could be applied at $0\nu\beta\beta$ energies and, in evaluating the performance of the network on the data domain, uses events outside the energy range within which we intend to make predictions. Namely, before applying the CNN to the peak itself, we evaluate the performance on the peak sidebands (see Fig.~\ref{fig.spectrum}), where the sample composition is known, and we expect the CNN predictions to be similar in data and MC. The underlying assumption is that the domain shift between MC and data is not correlated with the type of event, i.e. we expect that if a network is robust to MC/data differences on sidebands, it will be robust to MC/data differences in the peak region as well. In~\cite{Ferrario:2019} it was shown that the track length difference between data and MC is consistent across a wide energy range, giving us confidence that the differences are indeed coming from the detector simulation and reconstruction (which should have the same effect on both signal and background events), rather than incorrectly simulated physical processes, justifying the sidebands-testing approach.

\subsection{Network architecture}
After the initial proposal of a deep convolutional architecture for image recognition~\cite{Krizhevsky2017ImageNetCW}, all modern architectures are a combination of a set of basic layers: convolutional, downsampling (pooling) and fully connected (dense) layers. They are described as follows:
\begin{enumerate}
	\itemsep0em 
	\item[\textbullet] Convolution is performed by linearly applying \emph{filters}, which are tensors of learnable parameters whose dimensions correspond to the spatial dimensions of the input image, and stacking the output of each filter. For example, in the case of a two-dimensional colored image, a typical filter would have size $5\times5\times3$ (the last dimension corresponds to the \emph{feature}\footnote{In machine learning language, features are sets of numbers extracted from input data that are relevant to the task being learned.} dimension, which in the case of colored images represents the three RGB channels) and would slide\footnote{The step of the sliding is called \emph{stride} and the usual values are 1-2.} over the width and height of the input image, at each step applying a dot product over the third dimension and producing a 2D output. The process is repeated \emph{number-of-filters} times and the 2D outputs are stacked into a final layer output.
	\item[\textbullet] Pooling layers are used to reduce the spatial dimensions of the intermediate outputs, and usually a maximum or average value of the features inside a small window (typical size is 2) is used.
	\item[\textbullet] Fully connected layers, the basic layer of shallow neural networks, perform a matrix multiplication of a 1D input vector with a matrix of parameters. 
\end{enumerate} 
Other standard layers include dropout layers~\cite{10.5555/2627435.2670313}, where in each forward pass some number of features is randomly dropped during the training to ensure that the prediction is not dominated by some specific features, and normalization layers, in which the outputs of the intermediate layers are rescaled to avoid numerical instabilities. The most commonly employed normalization layer is the batch normalization layer~\cite{ioffe2015batch}.

A nonlinear activation function is applied\footnote{Without nonlinear activations, the whole network would simply collapse into a linear model.} to the output of each convolutional and dense layer. The choices of the activation functions are many, but the standard ones are the rectified linear unit (ReLU) function for intermediate layers, and the Softmax function for the final output in the case of classification. Softmax is used to map the input vectors to the range $[0, 1]$, hence allowing the outputs of the network to be interpreted as probabilities.\footnote{The predictions of deep networks should be scaled to truly represent probabilities (in the frequentist sense)~\cite{Guo2017OnCO}.} The number and order of layers and the number of filters and their sizes should be adapted considering the complexity of the problem, the size of the training dataset and the available computational resources.

\begin{figure}[htp!]
	\subfloat[ResNet architecture]{
		\begin{minipage}{0.5\textwidth}
			\centering
			\includegraphics[width=0.5\textwidth]{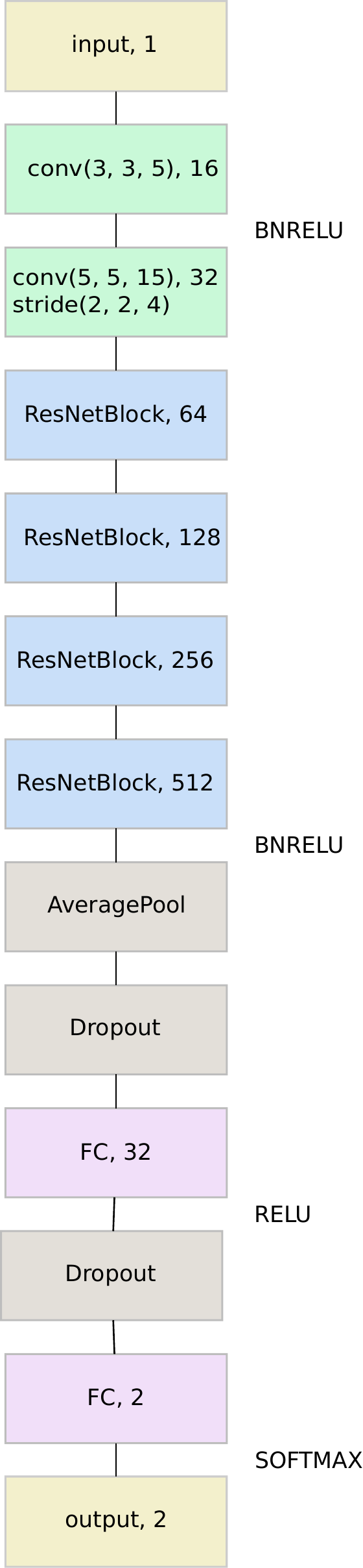}
	\end{minipage}}
	\subfloat[ResNet pre-activated block]{
		\begin{minipage}{0.5\textwidth}
			\centering
			\includegraphics[width=0.8\textwidth]{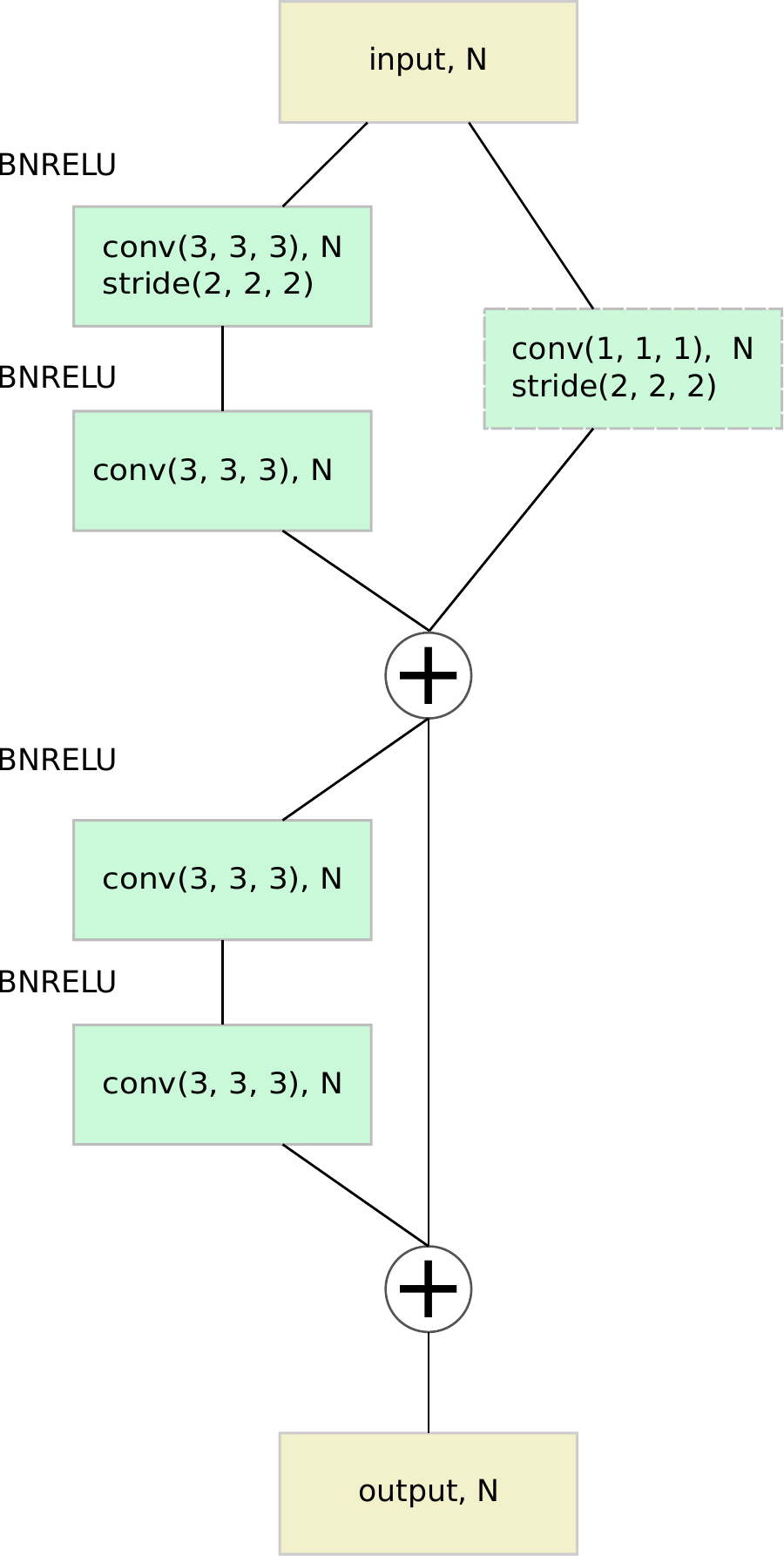}
	\end{minipage}}
	\caption{a) Summary of the neural network architecture used in this analysis, with b) details of each ResNetBlock architecture.
		The following notation is used: conv(fx, fy, fz),N represents a Convolutional 3D layer with filter size (fx, fy, fz) and N filters, with stride(sx, sy, sz) shown in the same cell if stride$>$1 is applied. FC, N represents a fully connected layer with N output neurons. AveragePool is applied after the last ResNetBlock and Dropout with p=0.5 is applied before each FC layer.  Nonlinear activation throughout the network is ReLU, except for the final activation which is Softmax. BatchNorm normalization layer is applied before ReLU activation for all but the last ReLU layer.} \label{fig.network}
\end{figure}

For the particular choice of network architecture in this study, we followed state-of-the-art practices in computer vision and deep neural networks, with special adaptations for the nature of our data.  The most common building block for image analyzing networks is the \emph{residual block} pioneered in \cite{He:2015}, with our implementation pictured in Fig.~\ref{fig.network} (b).  This building block uses convolutional filters to learn a modification to its input layer rather than a direct transformation: $R(x) = x + f(x)$, where the residual block $R$ is learning small changes ($f$) to an image.  The network used here consisted of two initial convolutional layers, and a set of pre-activated residual block layers \cite{He2016IdentityMI} (a particular variant of residual layers) followed by two consecutive dense layers with a dropout layer before each, as seen in Fig.~\ref{fig.network} (a). The nonlinear activation employed is the standard 
ReLU preceded by a batch normalization layer in all but the fully connected layers.\footnote{Batch normalization layers mixed with Dropout layers can lead to numerical instabilities~\cite{Li2019UnderstandingTD}.}

A major modification to our network from the residual networks in \cite{He:2015} is the use of three-dimensional convolutions as opposed to two-dimensional convolutions.  Additionally, we employ fewer layers since our classification task with two output classes is simpler than the thousand-class challenges confronted by ImageNet~\cite{deng2009imagenet}, for which residual networks were originally developed.  

In our network, the input dimensions were $40\times40\times110$ with each input corresponding to one voxel, therefore covering a volume of $40\times40\times55$~cm$^3$, essentially the entire active volume of the detector. The data are locally dense but globally sparse, as is common in high-energy physics and nuclear physics.  In this dataset, the number of voxels with activity is typically $\mathcal{O}(100)$, compared to 176,000 voxel locations in the input space (0.05\% pixel occupancy).

While traditional dense 3D convolutional neural networks will work on this dataset, we employed the Submanifold Sparse Convolutional Networks (SCN) framework \cite{Graham2017SubmanifoldSC}, implemented in PyTorch. SCN is highly suitable for sparse input data, making the linear algebra far more computationally efficient than with non-sparse techniques. 

Such networks have already been used in high-energy physics analysis \cite{Domin2019ScalableDC} and the main advantage of these types of network is that they occupy less memory and allow for larger input volumes and/or larger batch sizes,\footnote{Batch size is the number of images processed in one forward pass through the network. Larger batch sizes allow for faster training.} as well as significantly faster training times. All of the results shown here were obtained using this framework, but we obtained similar results using the standard implementation of dense convolutions in Keras/TensorFlow.

We note, in particular, that the dataset used here from NEXT-White allows for the use of dense convolutional neural networks, but this will not be true in future detectors. Applications of this technology at larger scales, including NEXT-100 and beyond, will lead to a total number of voxels that scales as the volume of the detector, growing with a cubic power.  Current and future AI accelerator hardware (GPUs, for example) will not be able to process the dense volumes of a ton-scale detector.  However, the number of active voxels for an interesting physics event will likely stay the same as in NEXT-White, or increase modestly with improved resolution.  Therefore, sparse convolutional networks will be scalable to even the largest high pressure TPCs in the NEXT program.

The main result of this paper is the achievement of reliable results from CNN-based tools that operate on real data, rather than working towards small gains in improvement on simulation with a neural architecture search.  As will be shown later, the training has to be constrained to prevent domain overfitting, hence some improvement on simulated data could be achieved with different architectures but would not necessarily be reflected in experimental data.

\subsection{Training procedure}\label{training_proc}
Optimizing learnable parameters of convolutional and fully connected layers to achieve the desired predictions is called training, and is done by minimizing the disagreement (\emph{loss}) between the true labels and the predicted ones.\footnote{This is a general paradigm behind \emph{supervised} machine learning.} The most common loss function employed in classification problems with neural networks is the cross entropy loss $H(p, q) = -\sum_{x}p(x)\log q(x)$, where $p(x)$ are the true labels and $q(x)$ the predicted ones, calculated over all input images of one batch. 

\begin{figure}[htp!]
	\centering
	\includegraphics[width= 0.85\textwidth]{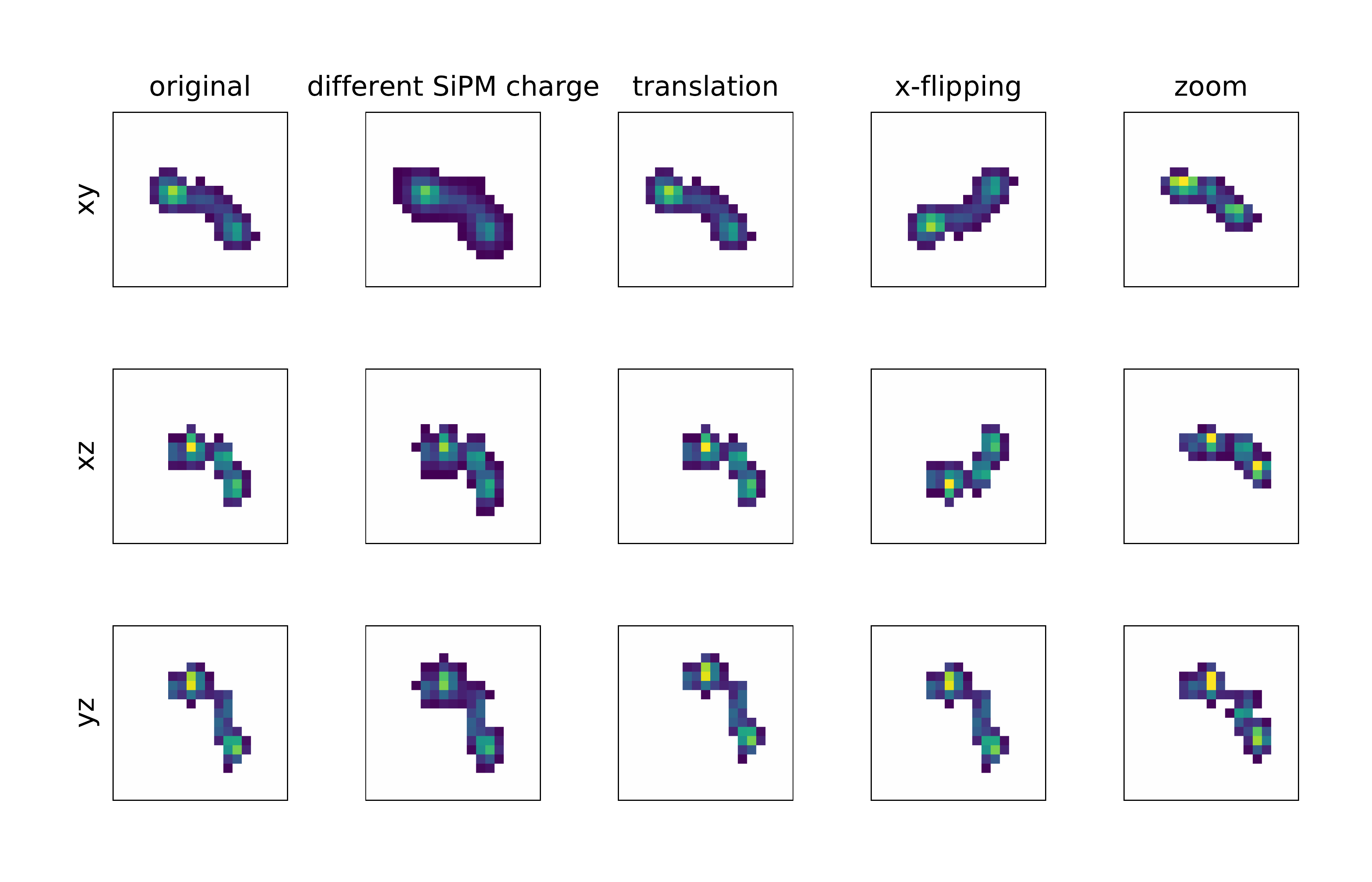}
	\caption{Example of on-the-fly data augmentation used during training on a selected signal event, projected on three planes for easier visualization.} \label{fig.imagetr}
\end{figure}

During the training, the performance of the network is tracked on the validation set, which is independent of the training set and is not used for network optimization. In this study, a total of about 500k simulated fiducial events were used as a training set, of which 200k were signal events, and an additional $\sim$ 30k events were used as a validation sample with similar signal proportion. A batch size of 1024 was chosen, and the cross entropy loss was weighted according to the signal-to-background ratio of the entire dataset. To avoid overfitting, in addition to the use of dropout, L2 weight regularization (a term in the loss function that penalizes high values of network parameters) and on-the-fly data augmentation\footnote{On-the-fly means that the augmentation is done during the code execution and the augmented dataset is not stored on the disk.}~\cite{Shorten2019ASO} were employed. The augmentation includes translations, dilation or ``zooming'' (scaling all 3 axes independently), flipping in x and y, and varying SiPM charge cuts\footnote{For validation and testing, the SiPM cut was fixed at 20 photoelectrons, while in the augmentation it was allowed to vary $\pm 10$ around this value.} as detailed in Fig.~\ref{fig.imagetr}. We note that augmentation procedures used here are explicitly designed to be ``label preserving'' in that they do not change the single or double-blob nature of events, but do reduce the significance of differences in data/simulation.

 As noted in section~\ref{ss.data_prep}, since CNNs are highly nonlinear models, their application outside the training domain cannot be assumed to be reliable, and before applying the network to events in the peak we compare extracted features of MC and data events on the sidebands. It is common to consider convolutional layers as feature extractors (each one extracting higher level features), and consecutive dense layers as a classifier. The paradigm of features extractors followed by a classifier is not unique to machine learning approaches. For example, the previous analysis of NEXT event topology used graph methods to determine connected tracks and extract track properties such as length, endpoints and the energies around them (so-called \textit{blobs}), which can be considered as a feature extractor. A classifier in that case was a cut on the obtained blob energies. Later on, the matching of MC and data features was done by rescaling MC features (for further explanation and justification of the procedure we refer the reader to~\cite{Ferrario:2019}). 
 
 The features in CNN models do not necessarily have intuitive meaning and should rather be considered as a dimensionality reduction that preserves relevant information from the input image needed for the classification task. Hence, the choice of the layer to be considered as the last one of the feature extractor is somewhat arbitrary. We chose the output of the AveragePool layer (Fig. \ref{fig.network}) as a representative feature vector, resulting in 512-dimension features. Since the features are not intuitive, any rescaling would not be justified and we are limited to simply tracking how MC features differ from those of data on the sidebands where the sample composition is known, and hence we expect similar distributions if the network is indeed robust to MC/data differences. To quantify similarity between multidimensional distributions we applied a two sample test; a test to determine whether independent random samples of $R^d$-valued random vectors are drawn from the same underlying distribution. We chose energy test statistics~\cite{Szekely04, Szekely13}, but the qualitative result should not depend on the choice of the metrics. The energy distance between two sets $A, B$ is given by
\begin{equation} \label{eq.ED}
\epsilon_{A, B} ={\frac {2}{nm}}\sum _{i=1}^{n}\sum _{j=1}^{m}\|x_{i}-y_{j}\| - {\frac {1}{n^{2}}}\sum _{i=1}^{n}\sum _{j=1}^{n}\|x_{i}-x_{j}\| - {\frac {1}{m^{2}}}\sum _{i=1}^{m}\sum _{j=1}^{m}\|y_{i}-y_{j}\| 
\end{equation}
where $x_i, y_i$ are $n, m$ samples drawn from the two sets. In~\cite{Szekely13}, it was proven that this quantity is non-negative and equal to zero only if $x_i$ and $y_i$ are identically distributed, hence the energy distance is indeed a metric. The p-value, or probability of observing an equal or more extreme value than the measured value, for rejecting the null hypothesis (in this case, rejecting the hypothesis that MC and data features follow the same underlying distribution) can be calculated via the permutation test~\cite{fisher1935design}. Namely, the nominal energy distance is computed, and the $x_i$ and $y_j$ are then divided into many (1000 in our case) possible arrangements of two groups of size $n$ and $m$. The energy distance is computed again for each of these arrangements, each of which corresponds to one permutation. The p-value is given by the fraction of permutations in which the energy distance was larger than the nominal one.

\begin{figure}[htp!]
	   \centering
	   \includegraphics[width=0.49\textwidth]{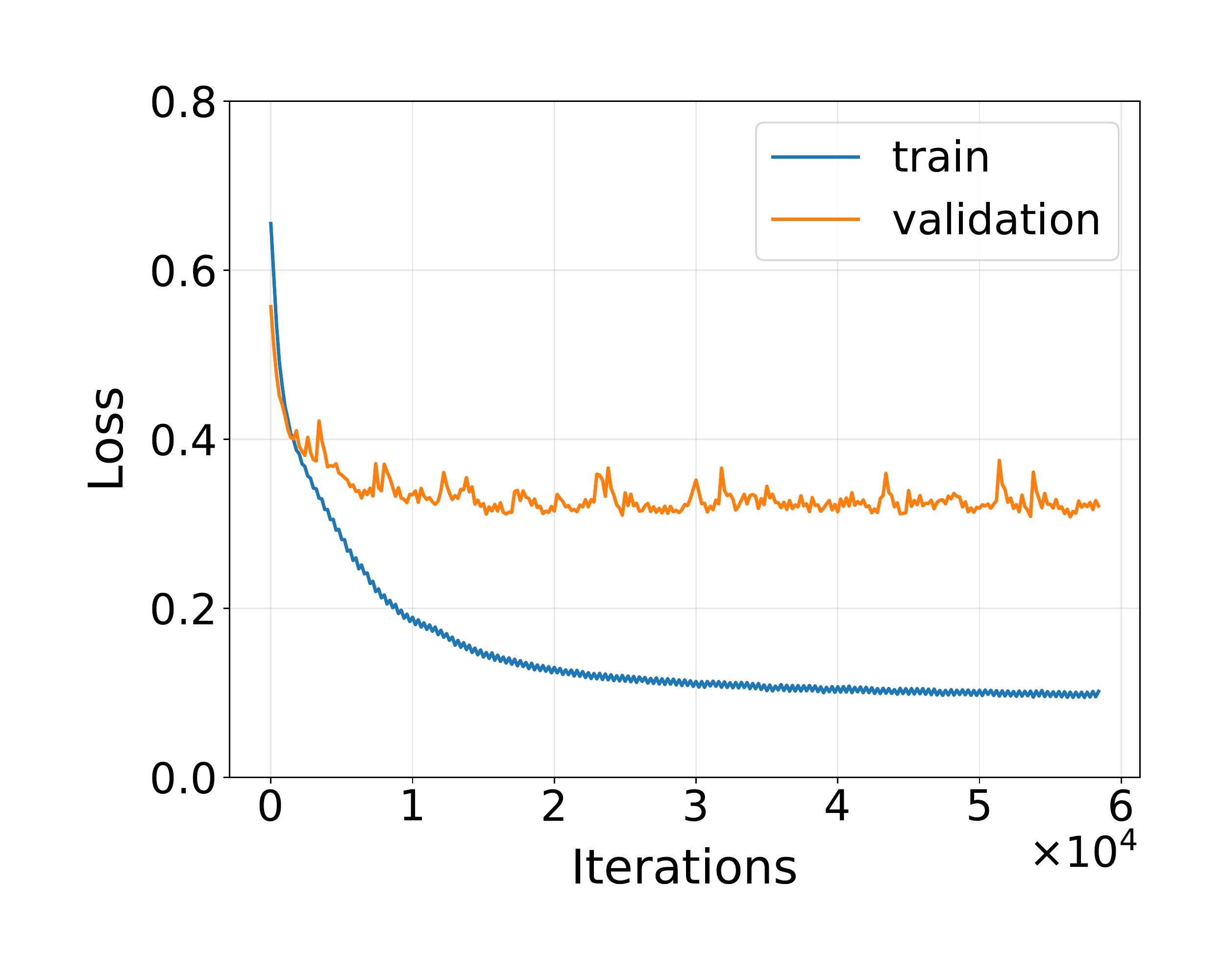} \includegraphics[width=0.49\textwidth]{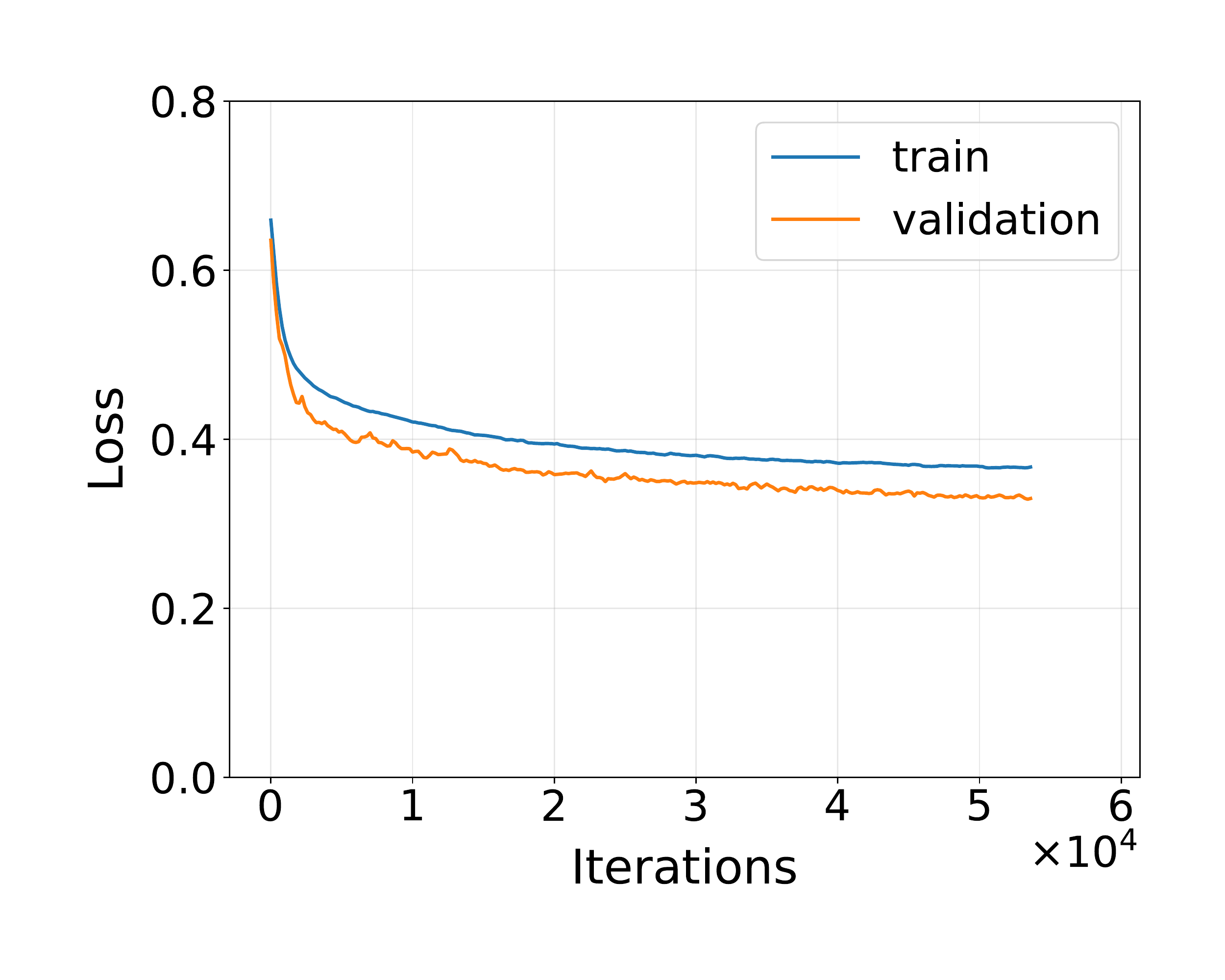}
	\caption{Training and validation losses without (left) and with (right) the application of data augmentation to the training set. Overfitting in the left-hand plot is visible only after 1000 iterations.  As the augmentation procedure is only relevant to the training phase, it was not applied to the validation set. The ability of the network to make correct predictions is improved for events unaltered by data augmentation, which explains why the loss is higher for the training set than for the validation set in the right-hand plot.} \label{fig.training_loss}
\end{figure}

\begin{figure}[htp!]
	   \centering
	   \includegraphics[width=0.49\textwidth]{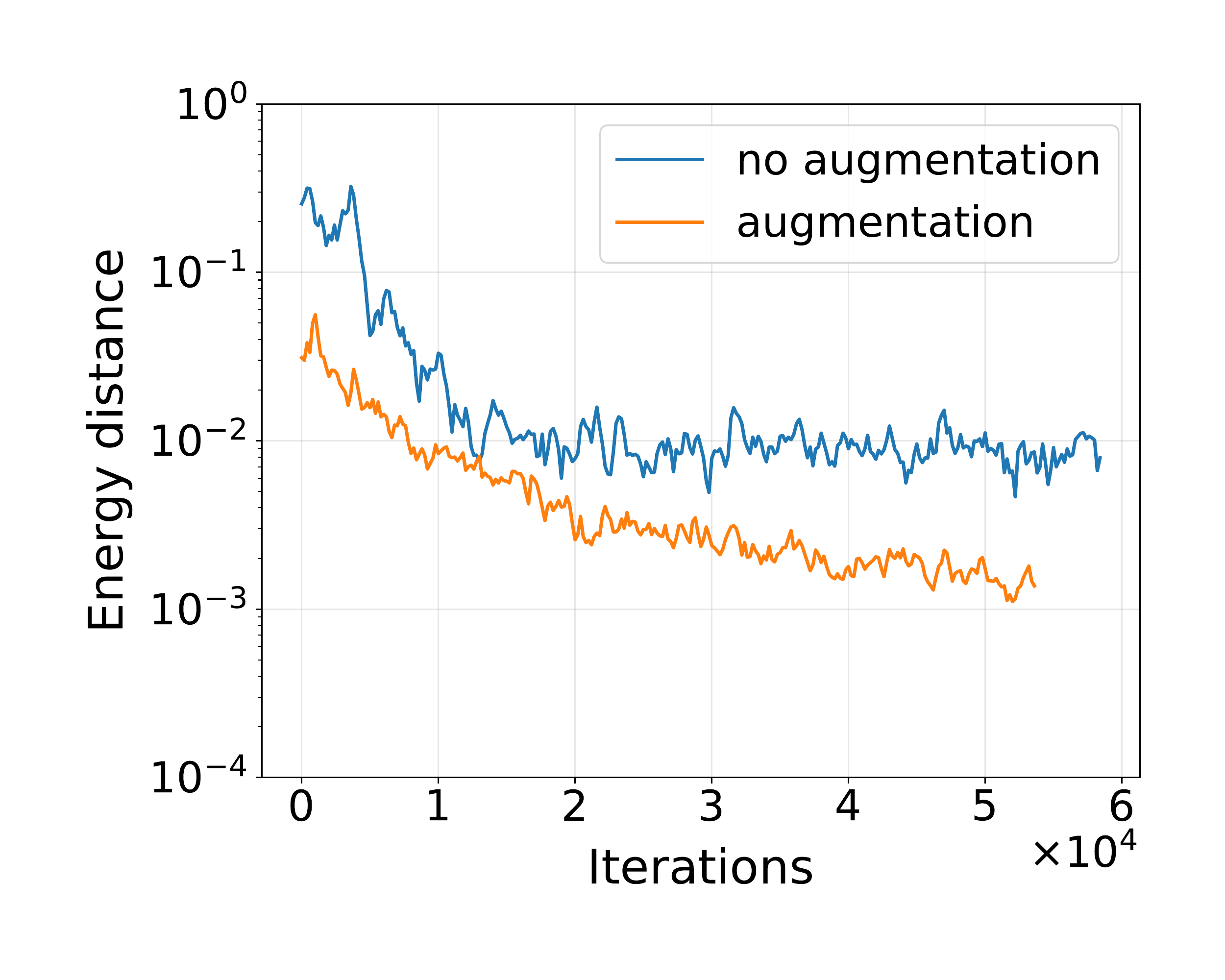} \includegraphics[width=0.49\textwidth]{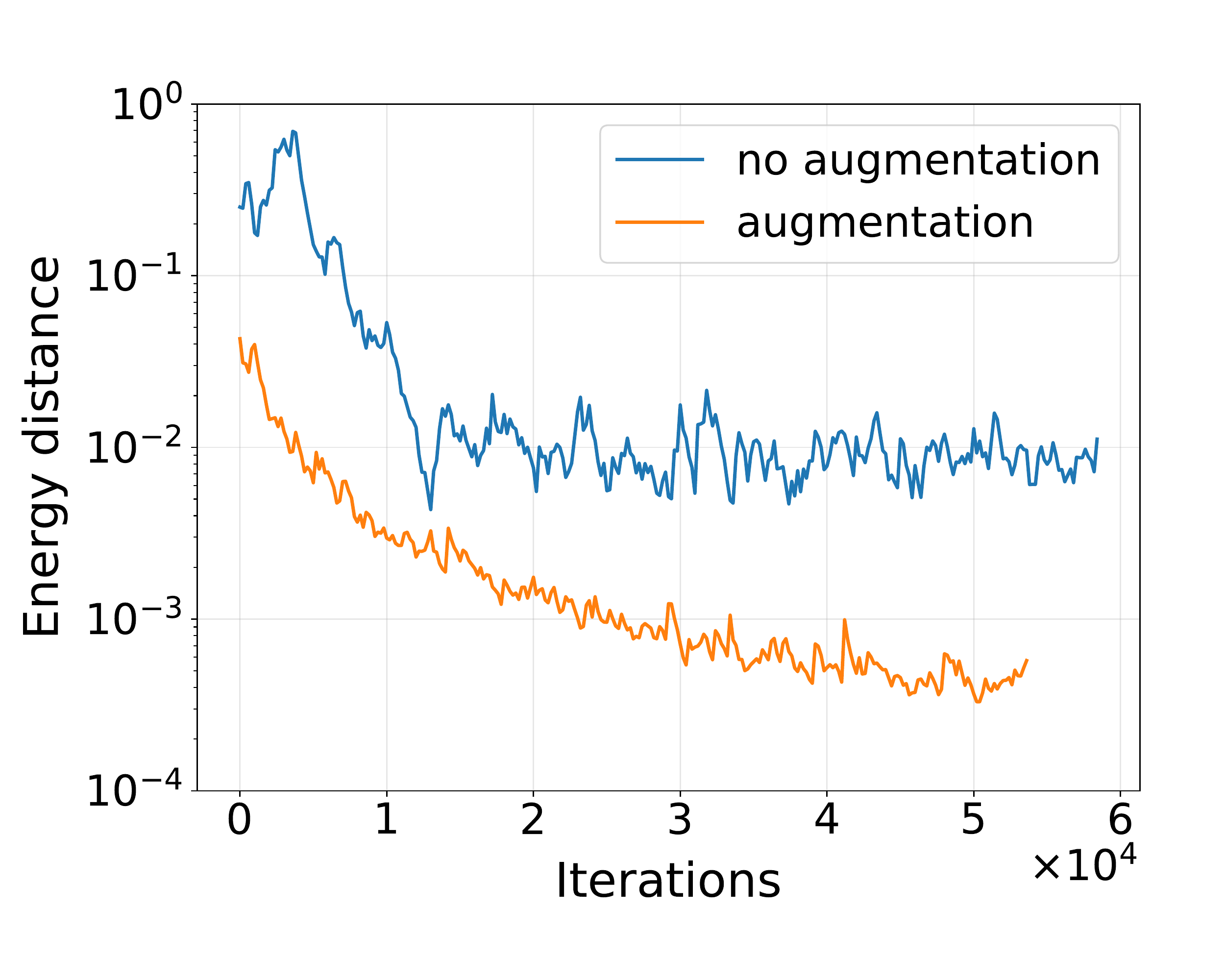}
	\caption{Energy distance between data and MC features during the training on the left sideband (left) and right sideband (right) for training with and without the augmentation. The corresponding p-value for the chosen model with augmentation at the chosen iteration step was $\sim$~0.1~(0.2) for the left~(right) sideband.} \label{fig.energy_loss}

\end{figure}

The training and validation losses are given in Fig.~\ref{fig.training_loss} for the networks trained with and without data augmentation. The overfitting apparent in the case of training without augmentation is prompt and is manifested in the divergence of the validation and test losses, meaning that the network is beginning to memorize the training dataset and is not generalizing well. In Fig.~\ref{fig.energy_loss} we show that the data augmentation also reduces the data/MC features distribution distance (eq.~\ref{eq.ED}), giving us more confidence that the performance on data will be similar to the performance on MC. As the distances are always calculated on MC and data events directly (without applying any data augmentation transformations), this technique does not directly correct MC but rather makes the model more robust to the data/MC differences. The final model is chosen by varying regularization parameters and selecting the training iteration step that gives minimal classification loss on the MC validation sample, ensuring that the corresponding p-value of energy test statistics is not larger than 5\%.

\subsection{Evaluation on data}\label{evaluation_data}
In an ideal test of the trained network, we would have a data sample of only $e^{+}e^{-}$ events at the energy of interest acquired from our detector, and another sample of single-electron events at the same energy. However, as we will always have background events, in particular due to Compton scattering of the high-energy gamma rays used in producing the $e^{+}e^{-}$ events with the topology of interest, an exactly-labeled test set of detector data is impossible. Therefore we make an assumption about the characteristics of the energy spectrum near the energy of interest and attempt to extract the number of signal and background events present, following the procedure explained in~\cite{Ferrario:2019}.

First, we select only fiducial events passing a single-track cut as explained in section~\ref{ss.data_prep}. Note that the single-track cut was not applied to the training set, but we do apply it to the test set to allow for exact comparison with the previous analysis. We then assume that the signal events produce a Gaussian peak (as indeed would be the case for events occurring at a precise energy), and that the background, consisting of Compton electrons, in the region of the peak can be characterized by an exponential distribution. The peak energy region is fixed to 1.570--1.615~MeV (as in~\cite{Ferrario:2019}), a region that contains more than 99.5\% of the Gaussian peak for both data and MC. Then, we apply an unbinned fit of the sum of two curves (Gaussian + exponential) to the full energy spectrum in the larger energy range\footnote{Note the slightly narrower energy range compared to the one specified in section~\ref{ss.data_prep} which is used to pre-select MC events based on true energy deposited in the detector. This choice is made to remove artificial disturbances at the end points of the selected energy range.} 1.45--1.75~MeV in order to keep the fits stable, obtaining the parameters defining the two curves. Integrating over theoretical Gaussian and exponential curves in the peak energy range gives us the estimate of the initial number of signal events $s_0$ (from the Gaussian) and the initial number of background events $b_0$ (from the exponential). This procedure is then repeated using the spectra obtained from events with network classification greater than a varying threshold, in each case obtaining the number of accepted signal events $s$ and accepted background events $b$.

\begin{figure}[!htb]
	\centering
	\textbf{Monte Carlo}
	\includegraphics[width= 1.0\textwidth]{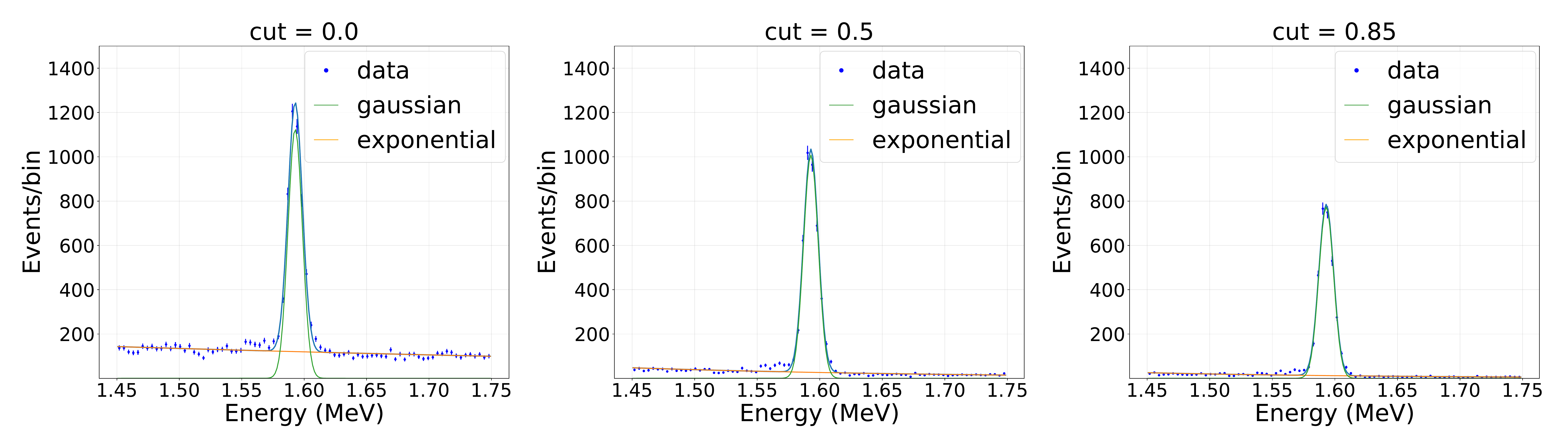}
	\textbf{Data}
	\includegraphics[width= 1.0\textwidth]{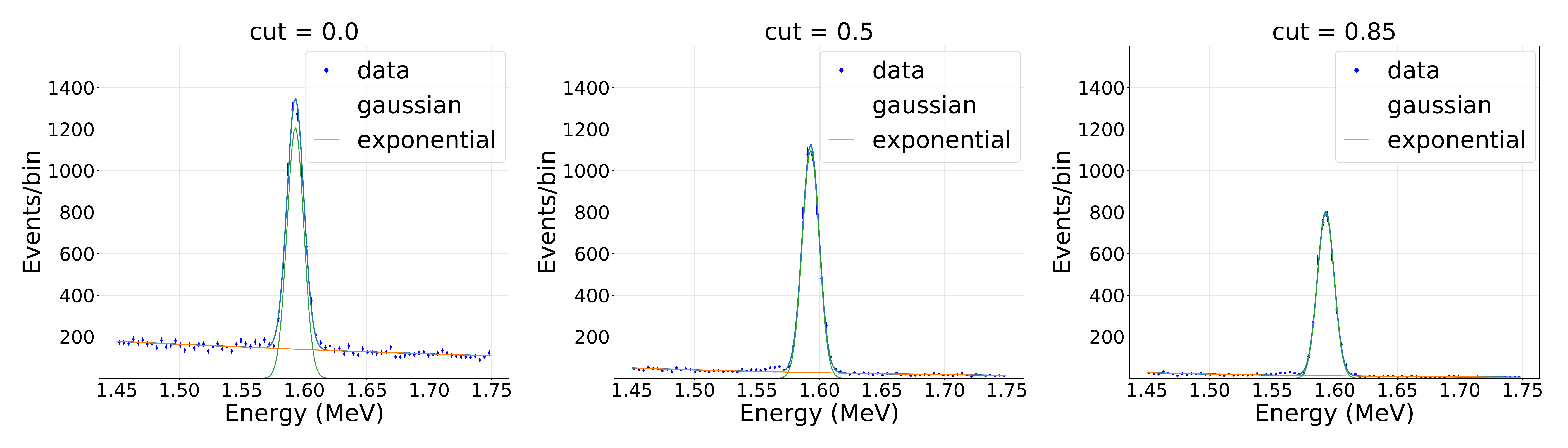}
	\caption{Fits of the energy spectra of Monte Carlo simulation (top) and data (bottom) near the double-escape peak of $^{208}$Tl at 1592~keV (see text for details). The fits are shown using events passing a neural network classification cut of 0.0 (to obtain the total number of signal and background events), 0.5 (the cut usually used in binary classification problems) and 0.85 (the cut that was found to yield the optimal f.o.m.)} \label{fig.fits}
\end{figure}

\begin{figure}[!htb]
	\centering
	\includegraphics[width= 0.48\textwidth]{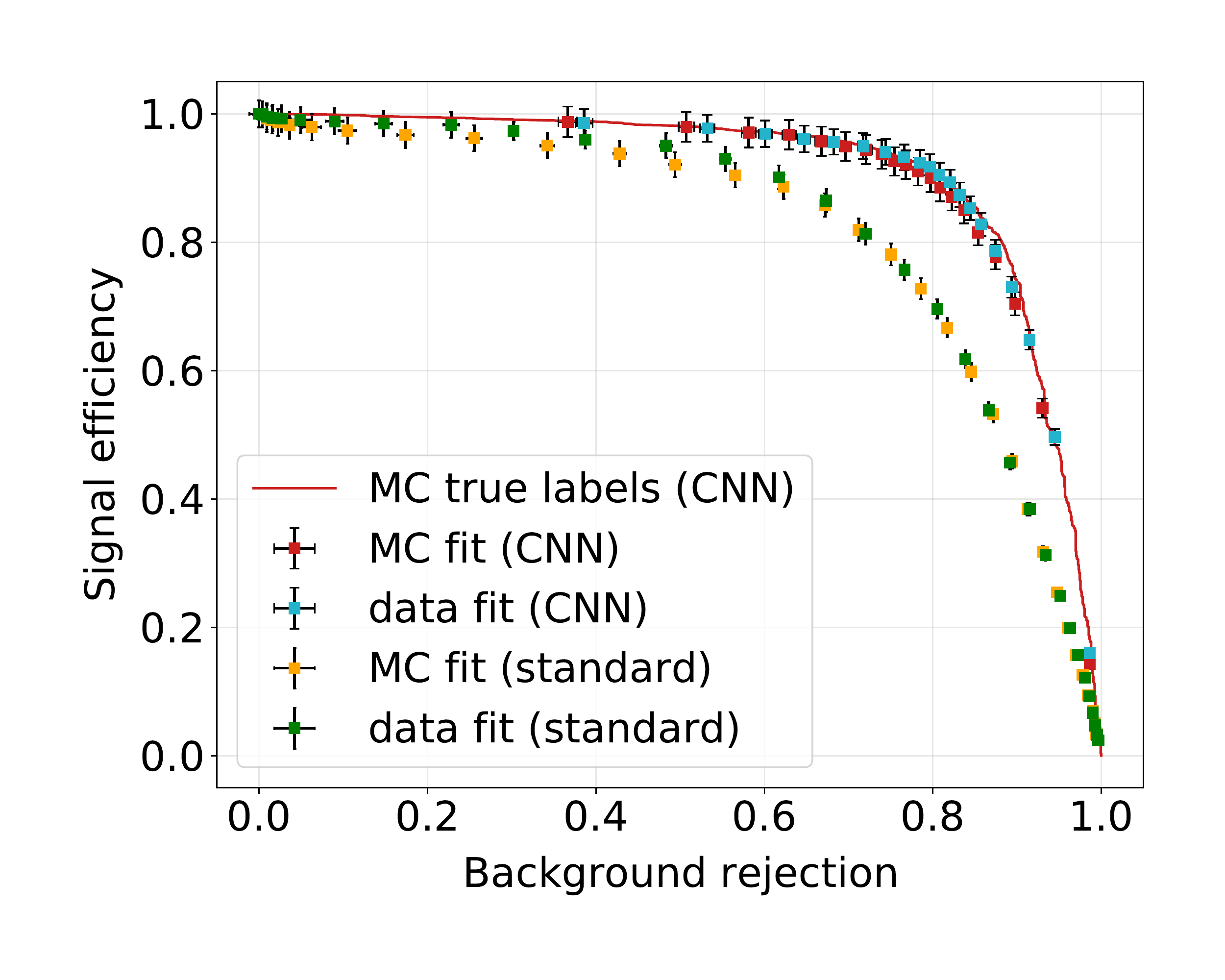}
	\includegraphics[width= 0.48\textwidth]{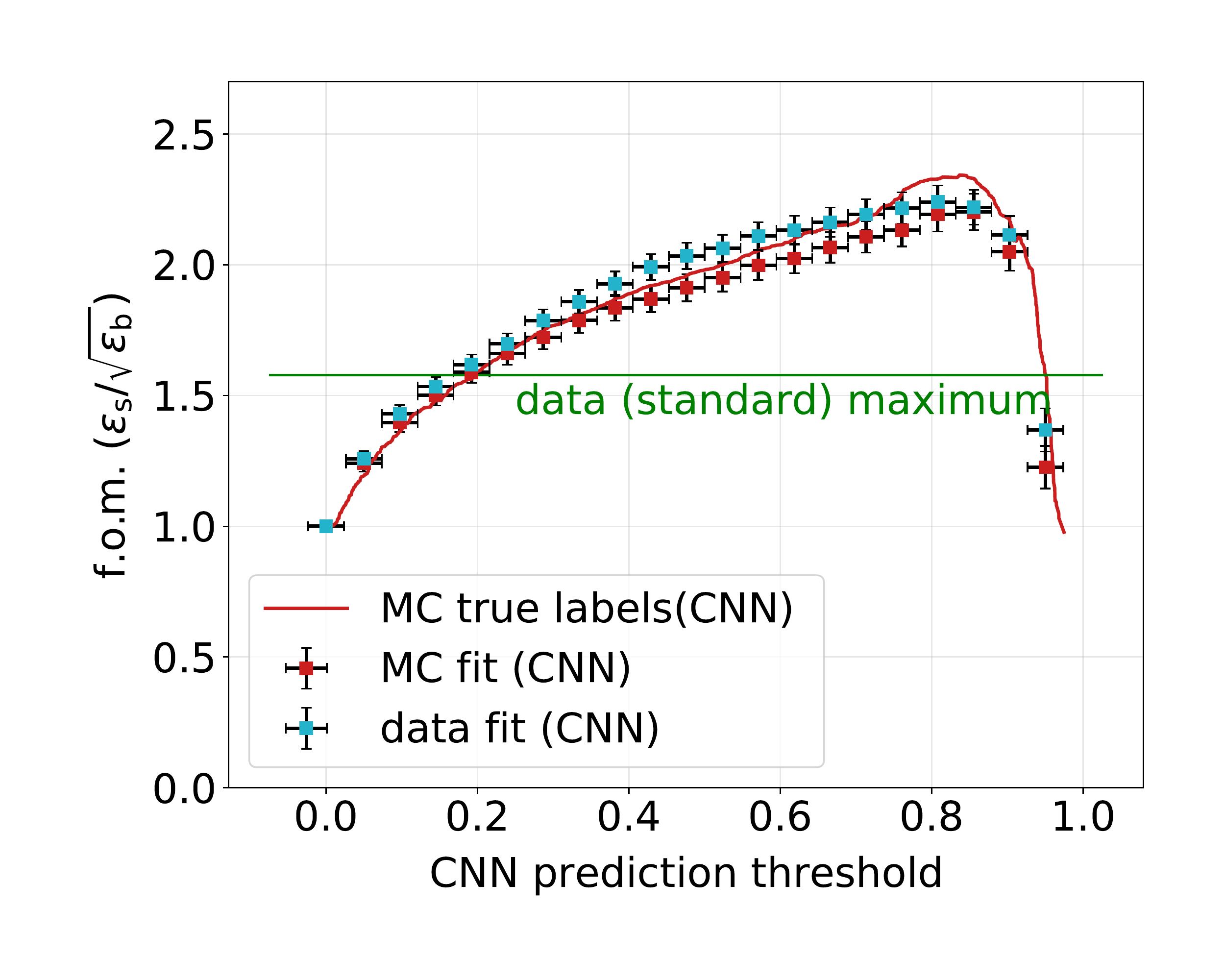}
	\caption{The signal acceptance vs. background rejection (left) and the figure of merit (right). The curves labeled ``fit'' are traced out by varying the neural network classification threshold and determining the fraction of accepted signal and rejected background using the fit procedure described in the text, while the MC true labels are obtained using MC labels as described in section \ref{ss.data_prep}. } \label{fig.sigvsbg}
\end{figure}

Figure~\ref{fig.fits} illustrates this fit procedure for three different threshold values on the CNN prediction output using data from NEXT-White and a set of Monte Carlo simulated events that were not present in the training set. Varying the classification threshold traces out a curve in the space of signal acceptance $s/s_0$ vs. background rejection $1 - b/b_0$ (see Fig.~\ref{fig.sigvsbg}). To obtain optimal sensitivity in a $0\nu\beta\beta$ search, one must maximize the ratio of accepted signal to the square root of the rejected background~\cite{Martin-Albo:2015rhw}, and therefore we also construct the figure of merit $F = s/\sqrt{b}$ for the various classification thresholds. We show for comparison the non-CNN-based result obtained in~\cite{Ferrario:2019}. In Monte Carlo, we find a maximum figure of merit of $F = 2.20$ with signal acceptance $s/s_0 = 0.70$ and background rejection $1-b/b_0 = 0.90$. In data, fixing the CNN cut to the one giving the best Monte Carlo figure of merit, we find $F = 2.21$, with signal acceptance $s/s_0 = 0.65$ and background rejection $1-b/b_0 = 0.91$. Compared to the previous result, the CNN analysis yields an improvement in the figure of merit of a factor $2.2/1.58 \sim 1.4$. Assuming that this factor stays the same at $0\nu\beta\beta$ energies, the improvement in the half-life measurement would be directly proportional to this factor, while the improvement in the measurement of $m_{\beta\beta}$ would be $\sqrt{1.4}\sim 1.16$. However, these factors should not be taken as exact since both traditional and CNN approaches depend on the detector specifics and reconstruction abilities, as well as the precision of the simulations.\footnote{In the case of no domain discrepancy between experimental and simulated data, the amount of regularization could be relaxed and the performance would improve.} The sensitivity predictions of the traditional event selection for the next two detectors planned can be found in~\cite{Martin-Albo:2015rhw, Adams:2020cye}.

We note that in Fig.~\ref{fig.sigvsbg} there is excellent agreement between data and simulation when comparing the signal efficiency in this analysis at a fixed background rejection, but there is still a minor disagreement between the figure of merit for simulation and data as a function of prediction threshold.  Several reasons account for this disagreement.  First, the data-augmentation technique extends the domain of applicability of the neural networks trained solely on simulated data, but it does not account for all possible differences between the data and Monte Carlo events. For example, any effect that would redistribute the energy along the track is not covered by the transformations we employ in data-augmentation. We anticipate that many of the effects contributing to data/simulation disagreement, such as the axial length effect mentioned in section~\ref{ss.evt_reco}, will be understood and resolved in the future and will bring these minor residual differences even closer together. A smaller EL TPC built from the original hardware of the NEXT-DEMO prototype~\cite{Alvarez:2012xda, Alvarez:2013gxa} is currently operational and will provide data that can be used to study these effects in more detail.

Second, the fit procedure error could account for some of the differences in the figure of merit plot. Namely, modeling the energy distribution as the sum of a Gaussian signal and exponential background does not adequately account for the long left tail visible in Fig.~\ref{fig.spectrum}. For low prediction threshold cuts, while the background acceptance is still high, these events are a minor effect, but as the threshold cut increases they become a larger portion of the left sideband during the fit procedure, leading to an underestimated signal efficiency when compared to the efficiency calculation on simulation obtained using the true underlying event type (the mismatch of red points and the continuous red line in Fig.~\ref{fig.sigvsbg}). A different ratio of signal inside the left sidebands between data and MC could lead to a different figure of merit.

	\section{Conclusions}\label{s.conclusions}
	We have demonstrated the first data-based evaluation of track classification in HPXe TPCs with neural networks. The results confirm the potential of the method demonstrated in previous simulation-based studies and show that neural networks trained using a detailed Monte Carlo can be employed to make predictions on real data. The present results show that the background contamination can be reduced to approximately 10\% while maintaining a signal efficiency of about 65\%. In fact, these results are likely to be conservative, as this demonstration was performed at an energy of 1592~keV, while at the same pressure, tracks with energy $Q_{\beta\beta}$ are longer, and therefore their topological features should be more pronounced. 

Furthermore, we have shown that, with the application of appropriate domain regularization techniques to the training set, our model performs similarly on detector data and simulation in the extraction of the signal events of interest.

	\acknowledgments
	This study used computing resources from Artemisa, co-funded by the European Union through the 2014-2020 FEDER Operative Programme of the Comunitat Valenciana, project IDIFEDER/2018/048. This research used resources of the Argonne Leadership Computing Facility, which is a DOE Office of Science User Facility supported under Contract DE-AC02-06CH11357.  The NEXT Collaboration acknowledges support from the following agencies and institutions: Xunta de Galicia (Centro singularde investigaci\'on de Galicia
accreditation 2019-2022), by European Union ERDF, and by the ``Mar\'ia de Maeztu`` Units
of Excellence program MDM-2016-0692 and the Spanish Research State Agency; the European Research Council (ERC) under the Advanced Grant 339787-NEXT; the European Union's Framework Programme for Research and Innovation Horizon 2020 (2014-2020) under the Grant Agreements No. 674896, 690575 and 740055; the Ministerio de Econom\'ia y Competitividad and the Ministerio de Ciencia, Innovaci\'on y Universidades of Spain under grants FIS2014-53371-C04, RTI2018-095979, the Severo Ochoa Program grants SEV-2014-0398 and CEX2018-000867-S; the GVA of Spain under grants PROMETEO/2016/120 and SEJI/2017/011; the Portuguese FCT under project PTDC/FIS-NUC/2525/2014 and under projects UID/FIS/04559/2020 to fund the activities of LIBPhys-UC; the U.S.\ Department of Energy under contracts number DE-AC02-07CH11359 (Fermi National Accelerator Laboratory), DE-FG02-13ER42020 (Texas A\&M) and DE-SC0019223 / DE-SC0019054 (University of Texas at Arlington); and the University of Texas at Arlington. DGD acknowledges Ramon y Cajal program (Spain) under contract number RYC-2015-18820. JM-A acknowledges support from Fundaci\'on Bancaria “la Caixa” (ID 100010434), grant code LCF/BQ/PI19/11690012. We also warmly acknowledge the Laboratori Nazionali del Gran Sasso (LNGS) and the Dark Side collaboration for their help with TPB coating of various parts of the NEXT-White TPC. Finally, we are grateful to the Laboratorio Subterr\'aneo de Canfranc for hosting and supporting the NEXT experiment.

	\bibliographystyle{JHEP}
	\bibliography{NextRefs}

\end{document}